
\input harvmac
\noblackbox

\def\l{\lambda}

\def\a{\alpha}
\def\b{\beta}
\def\d{\delta}
\def\t{\tau}
\def\s{\sigma}
\def\z{\zeta}

\def\r{\rho}

\def\ome{\Omega(E)}
\def\zbe{Z(\beta)}

\def\part{\partial}

\def\pa{\partial}
\def\o{\over}

\def\ha{{1\over2}}
\def\rar{\rightarrow}
\def\zbe{Z(\beta)}
\def\ie{ i.e.}
\def\hf{ {1\over 2}}
\def\cald{ {\cal D}}
\def\zb{ {\bar z}}
\def\qb{{\bar q}}
\def\Cb{{\bar C}}
\def\varthb{{\bar \vartheta}}
\def\cals{{\cal S}}
\def\calf{{\cal F}}

\def\cald{{\cal D}}
\def\frac#1#2{ {{#1}\over {#2}}}

\Title{\vbox{\baselineskip12pt\hbox{SUHEP-4241-522}}}
{{\vbox{\centerline{ Finite Temperature Strings$^*$}
\vskip2pt \centerline{ }
}}}
\bigskip
\centerline{ Mark\footnote{}{$^*$ Invited talk delivered
at the workshop ``String Quantum Gravity and Physics at the
Planck Energy Scale,'' Erice, June  21--28, 1992.} J. Bowick$^1$\footnote{}
{$^1$E-mail: Bowick@suhep.bitnet.}
}
\medskip
\centerline{ Physics Department }
\centerline{ Syracuse University}
\centerline{ Syracuse, NY 13244-1130, USA}
\bigskip
\centerline{\bf Abstract}
\medskip
In this talk a review of earlier work on finite temperature strings
was presented. Several topics were covered, including the canonical
and microcanonical ensemble of strings, the behavior of strings near the
Hagedorn temperature as well as speculations on the possible phases of
high temperature strings. The connection of the string ensemble and,
more generally, statistical systems with an exponentially growing density of
states with number theory was also discussed.
\noindent
\Date{ September 29, 1992}
\vfill \eject
%
%
%
%
\centerline{\bf 1. Introduction} \bigskip
The theory of fundamental strings has attracted a lot of interest in the past
decade as a possible candidate for a unified theory of all the known
interactions. Despite heroic efforts, crucial questions still remain
unanswered. For example, the basic degrees of freedom of this theory
are still not well understood. In particular near the Planck scale,
where we expect the stringy behavior to become more evident, it is
very poorly understood what is the correct configuration space and
which states dominate the dynamics of string theory at that scale.

One way to probe this region would be to study a very hot ensemble of
strings. There is, however, a difficulty as one tries to to heat the
ensemble past a limiting temperature $T_H = \b_H^{-1}$\nref\hag{
R. Hagedorn, {\sl Nuovo Cimento Suppl.} {\bf 3} (1965) 147.}
\nref\bowwi{
M. J. Bowick and L. C. R. Wijewardhana, {\sl Phys. Rev. Lett.} {\bf
54} (1985) 2485, and references therein.}\refs{\hag,\bowwi}, the famous
Hagedorn temperature. The number of states of string theory at a
specific energy level increases exponentially with the energy. As a
result, the partition function is infinite for $\b<\b_H$ due to
the competition between the entropy and the Boltzmann factor
$e^{-\b E}$. What happens when we try to increase the temperature past
this point is, despite much effort, merely a speculation.

Since the effective string tension vanishes as one approaches
$T_H$ one possibility is that pumping more energy into the ensemble
to increase its temperature merely results in the formation of longer
and ``wigglier'' strings by exciting higher oscillation modes
\nref\bowgi{ M. J. Bowick and S. B. Giddings, {\sl Nucl. Phys.} {\bf
B325} (1989) 631, and references therein.}\bowgi .
This might be related to a ``string uncertainty principle''
$$ \Delta x\approx {\hbar c\o E}+{G_NE\o g^2 c^5}\, ,$$
where as we try to probe smaller distances by scattering more
energetic strings, we obtain, after a point, poorer resolution due to
the formation of longer and longer strings.

Another suggestion is that the Hagedorn transition is really a first
or second order phase transition, where above $T_H$ we obtain a large
genus zero contribution to the free energy. In a string theory with
smooth world sheets we expect the genus zero contribution to the
partition function to be zero.
The reason is that the partition function is calculated by
toroidally compactifying the (euclidean) time direction with radius
$R=\b/2\pi$. Since the sphere is simply connected, it cannot wrap
around the time direction and it should give a trivial $\b$ dependence.
This suggests that smooth Riemann surfaces are not appropriate for
describing the high temperature phase of the string. The following
picture is suggested to hold \nref\sath{
B. Sathiapalan, {\sl Phys. Rev.} {\bf D35} (1987) 3277.}\nref\kog{
Ya. I. Kogan, {\sl JETP Lett.} {\bf 45} (1987) 709.}\nref\atwt{
J. Atick and E. Witten, {\sl Nucl. Phys.} {\bf B310} (1988) 291.}
\refs{\sath {--} \atwt}:
The modes $\varphi$ and $\varphi^*$ that
wind once around $S^1$ become tachyonic at $T_H$ and are supposed to
acquire nonzero expectation values $<\varphi>$ and $<\varphi^*>$
around which we should look for stable solutions (in fact due to the
dilaton the transition occurs at $T_{crit}<T_H$). Then the $\varphi$
vertex operator creates tiny holes on the world sheet that wind around
the time direction tearing the world sheet. This suggests that the
degrees of freedom in the high-$T$ phase are drastically reduced.
In fact the large $T$ behavior of the  free energy implies that there
are fewer fundamental gauge invariant degrees of freedom than any
known field theory that lives in the embedding space (for example for
the closed strings one finds that the number of degrees of freedom is
the same as that of a two dimensional field theory!).

At this point it is useful to consider the parallel with low energy QCD.
It has been claimed long ago that large $N$ QCD \nref\hft{
G. t' Hooft, {\sl Nucl. Phys.} {\bf B72} (1974) 461;
{\bf 75} (1974) 461.}\hft\ or the strong coupling
limit of QCD \nref\wls{ K. Wilson, {\sl Phys. Rev.} {\bf D10} (1974)
2445.}\wls\
can be formulated as a string theory (for a recent review see
\nref\plc{ J. Polchinski, {\sl ``Strings and QCD?''} U. Texas at
Austin Preprint, UTTG-16-92.}\plc\ and
references therein). In favor of this point of view is the early success of
the dual resonance models in describing the low energy meson resonances.
There are, however, difficulties with this picture. For example the
string in the relevant non-critical dimensions is not Lorentz
invariant in the light cone quantization and has one too many
oscillators because of the Liouville mode in the Polyakov quantization.

It is believed that $SU(N)$ QCD undergoes a first order deconfining
transition at some temperature $T_{dec}$. This can be the analogue of
the Hagedorn transition of string theory. We understand the high energy
regime because of asymptotic freedom. The basic degrees of freedom are
weekly interacting quarks and gluons. In the strong coupling limit
perturbative QCD is no longer a good description anymore, since the
perturbation series, even as an asymptotic series, is not a good
approximation.  Hadrons, mesons and glueballs are the
basic degrees of freedom and an effective string theory description in
terms of thin chromo-electric flux tubes might be possible.
The low temperature phase is described by smooth world sheets and
it is expected that such a description breaks down
in the high temperature regime;
it is the underlying
field theory that gives the basic degrees of freedom in that
case. In
string theory, however, a 26 or 10 dimensional field theory would be
inappropriate for the high temperature phase, since such a theory
would not have a good ultraviolet behavior.

The study of finite temperature strings also suggests some perhaps
deep connection between multiplicative number theory and statistical
systems with an exponentially growing density of states \nref\jul{
B. Julia, {\sl J. Phys. France } {\bf 50} (1989) 1371;
{\sl Statistical Theory of Numbers}, in {\sl Number Theory and
Physics}, eds. J. M. Luck, P. Moussa and M. Waldschmidt,
Proc. in Physics, Vol. {\bf 47} (Springer-Verlag, Berlin 1990)}\nref\spec{
D. Spector, {\sl Phys. Lett.} {\bf A140} (1989) 311;
{\sl Commun. Math. Phys.} {\bf 127} (1990) 239.}\nref\bowg{
M. J. Bowick, {\sl ``Quantum Statistical Mechanics and Multiplicative
Number Theory''} in {\sl ``Thermal Field Theories''}, Proc.
$2^{\rm nd}$ Workshop on Thermal Field Theories and their
Applications, Tsukuba, Japan, July 1990, eds. H. Ezawa, T. Arimitsu
and Y. Hashimoto, North-Holland, Amsterdam, 1991.}\nref\bowba{
I. Bakas and M. J. Bowick, {\sl J. Math. Phys} {\bf 32} (1991)
1881.}\refs{\jul {--} \bowba}. There exist
several examples, like the bosonic, fermionic or parafermionic Riemann gas
and the known string and conformal field
theories, where the partition function is related
to well studied number theoretic  multiplicative arithmetic functions
and modular forms. Understanding such a connection for physically
interesting string theories could be of great importance for calculating
and studying the behavior of relevant physical quantities.
The number theoretic methods developed are quite powerful
and can be used to calculate the partition function or
give a very good estimate of the asymptotic
behavior of the level densities of the above models. We should also not
underestimate the possible mathematical importance of these questions.

This presentation is organized as follows. In section two we introduce
the main relevant concepts. In section three we present the connection
between string theory at finite temperature and number theory.
We stress the relation between the level density
$p(N)$ of the bosonic string, the Dedekind eta function $\eta(\t)$
and Ramanujan's $\t$-function $\t(n)$. We explain how one can use
Hardy and Ramanujan's results to compute $p(N)$. In section four we
make this connection deeper by discussing the Riemann gas of bosons,
fermions or parafermions. In section five we discuss the
microcanonical and canonical ensemble of strings emphasizing the limit
of validity of the canonical ensemble near the Hagedorn temperature.
In section six we present the derivation of the asymptotic density of
states for the microcanonical ensemble and the effect of introducing
chemical potentials in order to impose conservation laws in the
computation of $\b_H$. In section seven we describe how we can obtain
the partition function with the string path integral on the torus.
In section eight we discuss the possible physical behavior of the
string ensemble near $\b_H$ and in section nine we discuss the results
of Kogan and Sathiapalan and Atick and Witten.
\bigskip
\centerline{\bf 2. Strings as a Statistical Mechanical System}\bigskip
Consider a $d$ dimensional string theory described by the embedding
$$
X:\Sigma\rar{\bf R}^{d-1,1}
$$
of the Riemann surface (complex curve) $\Sigma$ with coordinates
$(\s,\t)$ into the $d$ dimensional Minkowski space ${\bf R}^{d-1,1}$
with cartesian coordinates $X^{\mu}, \quad \mu=1, 2, \ldots, d$.
The classical free equations of motion
%
%
\eqn\eqmo{
(\partial _\sigma ^2-\partial_\tau ^2 )X^{\mu}= 0
}
and the boundary conditions, which for the open string
are $X^{\prime \mu }(\sigma ,\tau )=0$ for
$\sigma=0$ and $\pi $, determine the mode expansion
\eqn\modex{
x^\mu (\sigma ,\tau )=x_0^\mu +p^\mu \tau +i\sum_{n\neq 0}
{\alpha_n^\mu\o \sqrt{n}} e^{-in\tau }\cos n\sigma\, .
}
Upon quantization, the constraint
$({\rm L}_0-a)|\chi>=0$
on the physical states determines the spectrum of the
theory
\eqn\massq{
{1\o 2}M^2={\hat N} - a\, .
}
In the above formulas ${\rm L}_0$ is the zero mode of
the stress energy tensor, $a$ a normal ordering constant
and
$\hat N=
\sum_{n=1}^\infty n\alpha _{-n}\cdot \alpha _n=
\sum_{n=1}^\infty n\alpha _n^{\dagger}\cdot \alpha _n
$
is the level number operator.
The number of states at a particular level increases
very rapidly with $N$ and as we will soon see, it
increases exponentially with $N$ for large $N$. In
order to compute the number of states at level $N$, we
consider the generating function for the level
degeneracies
\eqn\levd{
\eqalign{
F(z)&\equiv \Tr\ z^{\hat N}\cr
    &=\sum_{N=0}^\infty d(N)z^N\cr
        }
}
where $d(N)$ is the number of mass eigenstates at level
$N$. We can compute $F(z)$ using
$
\hat N=\sum_{n=1}^\infty n\alpha _n^{\dagger}\cdot \alpha _n
$
\eqn\levco{
\eqalign{
F(z)&=\Tr\ z^{\sum\limits_1^{\infty _{}}n\alpha _n^{+}\cdot \alpha
_n}=\prod\limits_{n=1}^\infty \Tr\ z^{n\alpha
_n^{+}\cdot \alpha _n}\cr
    &=\left\{ \prod\limits_{n=1}^\infty \left(
     { 1\o 1-z^n}\right) \right\} ^{d-2}\cr
    &= |f(z)|^{d-2}\, ,\cr
        }
}
where \footnote{$^\dagger$}{The tachyon may be accounted for by
modifying this to $F(z)={1\o z}|f(z)| ^{d-2}$.}
$f(z)=\prod\limits_{n=1}^\infty
{1\o 1-z^n}=\sum\limits_{n=0}^\infty p(n)z^n$
is the classical partition function of Euler. It is
the generating function for the number of
unrestricted partitions of $N$ into positive integers.

\noindent{\it Proof:} Since
\eqn\levex{
f(z)=\prod\limits_{r=1}^\infty
(1+z^r+z^{2r}+z^{3r}+...)\, ,
}
a typical term $z^k$ in $f(z)$ is obtained by taking
one contribution from each factor $r=1,...,\infty $
such that
$
z^k=(z)^{k_1}(z^2)^{k_2}(z^3)^{k_3}...(z^n)^{k_n}
$
with
$
k=k_1+2k_2+3k_3+...+nk_n
$.
This yields a partition of $k$ into sets $\{k_i\}$.
\bigskip
\centerline{\bf 3. Singularities of $f(z)$ and Asymptotics
of $p(N)$}\bigskip
The function $f(z)$ has an essential singularity at all
{\it rational}
points of the unit circle $S^1$ i.e. the set of points
$\{e^{2\pi ip\o q}:p,q\in {\bf Z}\}\equiv\{z_{p,q}\}$.
The essential singularity arises
because $z_{p,q}^n=1$ for an infinite number of
integers $n=k\cdot q\, \forall k\in {\bf Z}.$
In order to estimate the asymptotic behavior of the
density of states we have to study the behavior of
$f(z)$ as $z\to 1^-$. A crude estimate gives
\eqn\calclv{
\eqalign{
\lim_{z\to 1^-} f(z)&=\lim_{z\to 1^-}
\prod\limits_{n=1}^\infty {1\o 1-z^n}=
\lim_{z\to 1^-}
\exp \left( -\sum\limits_{n=1}^\infty \ln
(1-z^n)\right)\cr
                    &=\lim_{z\to 1^-}
\exp \left(\sum\limits_{m,n=1}^\infty
{(z^n)^m\o m}\right) =
\lim_{z\to 1^-}
\exp \left( \sum\limits_{m=1}^\infty {z^m\o m(1-z^m)}
\right)\cr
                    &=\lim_{z\to 1^-}
\exp \left( \sum\limits_{m=1}^\infty
{1\o m^2}{1\o 1-z}\right)
        } .
}
This is the first appearance of exponential growth in the problem.
One can improve the above estimate by using the {\it modular}
properties of $f(z)$.
Writing $z$ as $e^{2\pi i\tau }$, the classical partition function
takes the form
\eqn\sfppa{
f(\tau )=\prod\limits_{n=1}^\infty {1\o 1-e^{2\pi i\tau }}\, ,
}
which is the
partition function of a single boson on the torus with modular
parameter $\t$.

This is the first connection we see with additive number theory
(combinatorics) and conformal field theory. The function $f(z)$ is
closely related to an important {\it modular form}, the Dedekind
eta function $\eta(\t)$ via
\eqn\etade{
\eta (\tau )=e^{i\pi \tau \o 12}
\prod\limits_{n=1}^\infty \left( 1-e^{2\pi in\tau }\right)
=z^{1\o 24}/ f(z)\, .
}
This function has many fascinating number theoretic properties.
For example
\eqn\etpro{
\eta ^{24}(\tau )={z\o f(z)^{24}}=z\left\{ \prod\limits_{n=1}^\infty
(1-z^n)\right\} ^{24}\equiv \sum\limits_{n=1}^\infty \tau (n)z^n
}
defines Ramanujan's $\t$-function $\t(n)$.

We list below some interesting properties of $\t(n)$
\item{1)}
 Ramanujan conjectured that $\tau (n)$ is a multiplicative function, i.e.
$$
\tau (mn)=\tau (m)\tau (n)\quad {\rm if}\quad (m,n)=1\, ,
$$
where $(m,n)$ is the greatest common divisor of $m$ and $n$.
This was proved by Mordell \nref\mord{L. J. Mordell, {\it
Proc. Cambridge Phil. Soc.} {\bf 19} (1917) 117.}\mord.
\item{2)}$\tau (m)\tau (n)=\sum\limits_{d\mid (m,n)}d^{11}
\tau ({mn\o d^2})$.
\item{3)}If $\tau (p)\equiv 0\ ({\rm mod}\ p)$
then $\tau (pn)\equiv 0\ ( {\rm mod}\ p)\quad\forall n$.
\item{4)} Dyson \nref\dys{F. J. Dyson {\sl Bull. Amer. Math. Soc.}
{\bf 78} (1972) 635.}\dys\ proved that
$$
\tau (n)=\sum\limits_{a,b,c,d,e}{
(a-b)(a-c)(a-d)(a-e)(b-c)(b-d)(b-e)(c-d)(c-e)(d-e)\o 1!2!3!4!}
$$
where $a,b,c,d,e$ satisfy
$$
a,b,c,d,e\equiv 1,2,3,4,5\ ({\rm mod}\ 5)\, ,
$$
$$
a+b+c+d+e=0\, ,
$$
and
$$
a^2+b^2+c^2+d^2+e^2=10n\, .
$$

The function $\eta(\t)$ is a modular form of weight $+\ha$ since
\eqn\modet{
\eta(-1/\tau )=(-i\tau )^{1/2}\eta (\tau )\, .
}
Using \modet\ we can map the region $z\to 1^-$, where $f(z)$ has
singularities, to the region $z'\to 0^+$, where $f(z)$ is
regular and close to $1$. Under $\t\to -1/\t$, $z=e^{2\pi
i\tau }\rightarrow z^{\prime }=e^{-2\pi i/\tau }$ and since
$ z'= \exp\left( {4\pi^2\o \ln z} \right)$,
we obtain
$$
f(z^{\prime })={(z^{\prime })^{1/24}\o \eta (-1/\tau )}=\left(
{-\ln z\o 2\pi }\right) ^{-1/2}z^{-1/24}f(z)
\ (z^{\prime })^{1/24}\ .
$$
Thus
\eqn\asyf{
\lim_{z\to 1^-} f(z)=
\lim_{z\to 1^-} \left( {-\ln z\o 2\pi} \right)^{\ha}
z^{1\o 24} \exp\left( {-\pi^2\o 6\ln z} \right)
}
Note that \asyf\ is a slightly improved estimate of the behavior
of $f(z)$ as $z\to 1^-$ as compared to \calclv .

To extract $p(N)$ from the above results we follow Hardy and
Ramanujan \nref\hr{ G. H. Hardy and S. Ramanujan,  {\sl Proc. London Math.
Soc.} {\bf 17} (1918) 75.}\hr\
who first applied the techniques of complex analysis to the
combinatoric problem of determining the asymptotic behavior of
the number of unrestricted partitions of an integer $N$.
\eqn\contpn{
p(N)={1\o 2\pi i}\oint dz\ {f(z)\o z^{N+1}}\, ,
}
where the contour is taken around the origin and within and close
to the unit circle. Then
\eqn\contlim{
p(N)={1\o 2\pi i}\oint dz\ \left( {-\ln z \o 2\pi }\right)
^{1/2}\exp \left[ {-\pi ^2\o 6\ln z}-(N+{23\o 24})\ln
z\right]\, .
}
A saddle point evaluation of $p(N)$ (see appendix) leads to the
result
\eqn\aspn{
p(N)\approx {1\o 4\sqrt{3}N}\exp \left( \pi \sqrt{2\o 3}\sqrt{N}%
\right)\, .
}
A better result for $p(N)$ was derived by Hardy and Ramanujan.
They showed that
\eqn\hrpn{
\eqalign{
p(N)&={1\o 2\pi \sqrt{2}}{d\o dN}\left( {e^{c\lambda _N}\o \lambda
_N}\right) +{(-1)^N\o 2\pi }{d\o dN}\left( {e^{ {1\o 2}c\lambda
_N}\o \lambda _N}\right)\cr
&\quad +{\sqrt{3}\o \pi \sqrt{2}}\cos
\left( {2\o 3}N\pi -{1\o 18}\pi
\right) {d\o dN}\left( {e^{ {1\o 3}c\lambda _N}\o \lambda _N}\right)
+\ldots\cr
        }
}
where $\l_N=\sqrt{N-{1\o 24}}$ and $c=\pi \sqrt{2\o 3}$. For
$N=100$ and $200$, summing 6 terms of this series gives
$p(100)=190,569,291.996$ and $p(200)=3,972,999,029,338.004$,
whereas the exact results are $p(100)=190,569,292$ and
$p(200)=3,972,999,029,388$ respectively \hr.
Later an exact analytical result was found by
Rademacher \nref\rade{H. Rademacher, {\sl Proc. London Math. Soc}
{\bf 43} (1937) 241.}\rade\ using a slight variant of the
Hardy-Ramanujan analysis
\eqn\radpn{
p(N)={1\o \pi \sqrt{2}}\sum\limits_{q=1}^\infty A_q(N)q^{1\o 2}
{d\o dN}\left( {\sinh \left( {c\lambda _N\o q}\right) \o \lambda _N}
\right)\, ,
}
where $A_q(N)$ is a certain sum of roots of unity.
\bigskip
\centerline{\bf 4. Quantum Statistical Mechanics and Multiplicative
Number Theory}\bigskip

Multiplicative number theory in the context of statistical systems
with exponentially  growing number of states has been considered first
by Julia and Spector\refs{\jul {--} \spec}.
Consider, instead of the usual Fock space of a
quantum field theory, a different labelling of the quantum states.
Place the creation operators $\{ a_i^\dagger \}$ in cardinal order and
associate the $i^{\rm th}$ prime number with the $i^{\rm th}$ creation
operator. Then label the state $\prod_i (a_i^\dagger)^{r_i}|0>$ by the
integer $N= \prod_i p_i^{r_i}$. Since the factorization of $N$ into
prime numbers is unique, this provides a basis for the Hilbert space
of states. This is known as the G\"odel numbering
\nref\godl{K. G\" odel, {\sl ``\"Uber Formal Unentscheidbare S\"atze
der Principia Mathematica und Verwandter Systeme I''}, {\sl
Monatshefte f\"ur Mathematik und Physik}, {\bf 38} (1931) 173.}\godl
. We are
arithmetizing this way a free quantum field theory the same way as the
formal calculus of propositions is arithmetized by associating the
$i^{\rm th}$  symbol in a proposition with the
$i^{\rm th}$ prime number and determining the corresponding
multiplicity by the symbol itself.

A simple example of a system with exponential growth of the number of
states is the Riemann gas. The partition function of such a system is
defined as
\eqn\nemp{
Z(\b)= \Tr \, e^{-\b H}\, ,
}
with energies $E_i= \ln\, p_i$. Then
\eqn\corip{
\eqalign{
Z(\b)&= \sum_{\{ r_i \} }\exp(-\b \sum_i r_i \ln\, p_i)\cr
     &= \sum_{N=1}^{\infty} {1\o N^{\b}} = \z(\b)\cr
        }
}
is the (ordinary) Riemann zeta function. The partition function
therefore, has a simple pole at $\b=1$. This can alternatively be seen
by computing the density of states
\eqn\zdns{
\eqalign{
\r(E) &= {1\o \Delta E} \r(N)\Delta(N)=\cr
      &= {1\o \ln\,\left({N+1 \o N} \right)} = N( 1+{1\o 2N}+\ldots)=\cr
      &= \exp (E) +{1\o 2}+ {\cal O}(e^{-E})\, .\cr
        }
}
We see that $\b_H=1$ corresponds to the simple pole of $\z(\b)$ at
$\b=1$.

Generally, multiplicative generating functions of the form (Dirichlet
series)
\eqn\dirser{
F(\b)= \sum_{n=1}^\infty {f(n)\o n^\b}
}
are of interest, where $f(n)$ is a general multiplicative arithmetic
function
on the natural numbers valued in some field (usually the field of real
or complex numbers). As in the case of the ordinary Riemann zeta
function, $F(\b)$ generally has a simple set of singularities but
a rich structure of complex zeroes.

The case $f(n)=\mu(n)$, where $\mu(n)$ is the M\"obius function, gives
the partition function of the fermionic analog of \nemp . This gives
$Z_F(\b)={\z(\b)\o \z(2\b)}$. Similarly the system of $k$-parafermions
is associated with \bowba\
\eqn\kmob{
f(n)\equiv \mu_k(n)=
\cases{ 1 &  if $n=\prod_i(p_i)^{r_i}$ with $0\le r_i\le k-1$;\cr
        0 &  otherwise.\cr
      }
}
It gives
\eqn\kpapa{
Z_k(\b)={\z(\b)\o \z(k\b)}\, .
}
Parafermions of order $2$ and $\infty$ correspond
to fermions and bosons respectively. Note that a representation of the
ordinary Riemann zeta function can be obtained by tensoring an
infinite set of parafermionic gases of order $k$ at successively lower
temperatures, such that
\eqn\infpfm{
\z(\b) \equiv Z_\infty(\b)= \prod_{n=0}^\infty Z_k(k^m\b)\, .
}
We hope that the above exposition has given a flavor of the important
connection between arithmetic gases and multiplicative number theory.
Much remains to be understood for the case of statistical systems that
arise from string theories and conformal field theories.
\bigskip
\centerline{\bf 5. The String Microcanonical and Canonical
Ensemble}\bigskip

We are now ready to compute the microcanonical density of states
for the string ensemble. Since the mass spectrum is given by
\eqn\msstw{
{1\o 4} M^2 = N - {\rm const}\, ,
}
the asymptotic form for $p(N)$ translates into the string density
of states
\eqn\denst{
\Omega(M)\sim M^{-\a} \exp{(\b M)}\, .
}
The microcanonical density of states $\ome$ is defined by
\eqn\densen{
\eqalign{
\ome &= \Tr\ \d(E-\tilde H)\cr
     &= \int {d\tilde \beta \o 2\pi }
       \Tr\ e^{i\tilde \beta (E-\tilde H)}\cr
     &=\int\limits_{\beta _0-i\infty }^{\beta _0+i\infty }
       {d\beta \o2\pi i}e^{\beta E}
       \Tr\ e^{-\beta \tilde H}\quad \cr
     &=\int\limits_{\beta _0-i\infty }^{\beta _0+i\infty }
       {d\beta \o2\pi i}e^{\beta E} Z(\b)\, ,
        }
}
where $\beta =i\tilde \beta$ and  $\b_0$ is chosen for the convergence of
$Z(\b)\equiv \Tr\ e^{-\beta \tilde H}$, the canonical partition
function. Inverting the Laplace transform we obtain
\eqn\partdf{
Z(\b) = \int_0^{\infty} dE\, \ome e^{-\b E}
}
In a typical statistical mechanical system with $N$ degrees of
freedom $\ome\sim E^N$ for large $E$. Thus the entropy $S=\ln
\ome\sim N\ln E$ is extensive. For large $N$, the integrand is
sharply peaked and the canonical partition function can be
evaluated by a saddle point approximation. Since
\eqn\sadpar{
Z(\b) = \int_0^{\infty} dE \exp{ \left(\ln\ome - \b E\right)}
}
the saddle point occurs at $E_0$ such that
\eqn\sadcon{
 {d\ln\Omega (E)\o dE}\bigg| _{E=E_0}=\beta ={N\o E_0}\, ,
}
i.e. $E_0={N\o \b}= N\,T$.
Thus fixing the temperature $T$ (canonical ensemble) is equivalent in the
thermodynamic limit to fixing $E$ in the microcanonical ensemble
at $N\, T$
and $T$ is really the average energy per degree of freedom.
The fluctuations are given by
\eqn\parflc{
\eqalign{
Z(\beta ) & = Z_0\int\limits_0^\infty dE\ \exp \left\{ {1\o 2}(E-E_0)^2
{\partial ^2\o \partial E^2}(S(E)-\beta E)\right\}\cr
          & = Z_0\int\limits_0^\infty dE\ \exp \left\{
              {1\o 2}(E-E_0)^2 {\partial ^2S(E)\o \partial
               E^2}\right\}\, .\cr
        }
}
This may be rewritten in terms of the microcanonical (formal)
specific  heat $C_V$. Since $S_M = \ln\ome$ and $T_M =\left(
{\pa E\o \pa S}\right)^{-1}$, we have
\eqn\cvmde{
{\pa^2 S\o\pa E^2}=-{1\o T^2}{dT\o dE}\bigg|_V=
-{1\o T^2}{1\o (C_V)_M}\, ,
}
where $C_V= {dE\o dT}$. Then
\eqn\cvm{
\left( C_V \right)_M = - {1\o T^2}\left( {\pa^2 S\o \pa
E^2}\right)^{-1} \, .
}
Thus \parflc\ becomes
\eqn\parsec{
Z(\b)= Z_0 \exp\left\{ -{\ha} (E-E_0)^2 {1\o T^2}\left( C_V
\right)_M^{-1} \right\}\, .
}
Convergence requires that $C_M>0$ in order that saddle point
approximation is valid.

For strings $\ome\sim E^{-\a}\exp{\b_M E}$ and as one approaches $T_H$
from below the integrand $\ome E^{-\b_M E}$ flattens out. Typically
$\a>1$ and $Z(\b)$ is well defined for $\b=\b_H$, but it diverges for
$\b>\b_H$. Fixing $T$ as $\b\to\b_H^+$ no longer corresponds to fixing
a precise $E$ and the canonical and microcanonical ensemble need not
agree. The entropy $S=\ln\ome\sim\b_H E- \a\ln E$ is no longer
extensive as well. For further discussion see section 7.

For free particles
$$
Z(\beta )=\prod\limits_{k_1b}\left( 1-e^{-\beta E_{k_1b}}\right)
^{-1}\prod\limits_{k_1f}\left( 1+e^{-\beta E_{k_1f}}\right)\, .
$$
Therefore
\eqn\lnpar{
\ln Z(\b) = \sum_{r=1}^\infty\left( {f_B
(\beta r)\o r}-(-1)^r{f_F(\beta r)\o r}\right)  \, ,
}
where the single-string partition functions $f_B $ and $f_F$ are
\eqn\spar{
\eqalign{
f_B&=\sum\limits_{k,b}e^{-\beta E_{k,b}}=\int\limits_0^\infty dE\ e^{-\beta
E}\omega _B(E)\cr
f_F&=\sum\limits_{k,f}e^{-\beta E_{k,f}}=\int\limits_0^\infty dE\ e^{-\beta
E}\omega _F(E)\cr
        }
}
Substituting \lnpar\ into \densen\ we obtain the multi-string density
of states from the single string density of states
\eqn\denbf{
\Omega (E)=\int\limits_{\beta _0-i\infty }^{\beta _0+i\infty }\ {d\beta
\o 2\pi i}e^{\beta E}\exp \left\{ \sum\limits_{r=1}^\infty \left( {
f_B(\beta r)\o r}-(-1)^r {f_F(\beta r)\o r}\right) \right\}\, .
}
The $r=1$ term gives the familiar Maxwell-Boltzmann expression
\eqn\boltz{
\Omega (E)=\sum\limits_{n=1}^\infty {1\o n!}\int \prod\limits_{i=1}^n\
dE_i\ \omega (E_i)\delta (E-\sum\limits_iE_i)\,
}
where
$
\omega (E)=\omega _B(E)+\omega _F(E)
$
\bigskip
\centerline{\bf 6. From the One-String Density of
States to the Multi-String }
\centerline{\bf Density of States for Toroidal
Compactification} \bigskip

Consider $d$-dimensional strings compactified on an internal manifold
of dimension $c$ that is for simplicity taken to be a torus with radii
$R_i$ with
$ i=1,2,\ldots, c$. In computing the full density of states, it
is important to incorporate the effects of
conservation laws \nref\deo{
N Deo, S. Jain and C. Tan, {\sl Phys. Lett.} {\bf B220}
(1989) 125.}\deo\ such
as conservation of momentum, winding number and, in the case of the
heterotic string, of $U(1)$ charges or windings/momenta in the
internal directions. For every conserved charge $Q_A$ we introduce a
chemical potential $\mu_A$ and compute
\eqn\mdefs{
\eqalign{
\Omega(E,\mu_A) &= tr\,\left[
e^{2\pi i \mu_A Q_A} \delta(E-H)\right]\cr Z(\beta,\mu_A) &=
tr\,\left[ e^{2\pi i \mu_A Q_A} e^{-\beta H}\right]\, .\cr } }
The condition that the total conserved charge be some fixed value
$q_A$ (typically $q_A=0$) is then enforced by multiplying
$\Omega(E,\mu_A)$ or $Z(\beta,\mu_A)$ by $\exp\{-2\pi i \mu_A q_A\}$
and integrating $\mu_A$ over the range $(-\hf,
\hf)$
\eqn\consds{
\Omega(E,q_A) = \int^\hf_{-\hf}d\mu_A\, \Omega(E,\mu_A)
e^{-2\pi i \mu_A q_A}\, .  }
The full density of states will be derived from the single-string
density of states as discussed in the previous section. In order to
compute the single string density of states, consider the energy of a
general string state
\eqn\compen{
E^2 =k^2+ \sum_{i=1}^c \left( {n_i^2\over R_i^2} + (2R_i)^2
m_i^2\right) +4n_L +4n_R\, .  }
Here $k$ is the momentum in the non-compact directions and $n_i$ and
$m_i$ are the momentum and winding quantum numbers, respectively. The
integers $n_L$ and $n_R$ label the internal level numbers describing
oscillatory mode excitations of the string.

As described in section 1, the number of left-moving (right-moving)
excitations at a given level is given by the generating function
$F(z)=\sum_\alpha z^{n_{L\alpha}}$ ($F(\zb ) = \sum_\alpha
\zb^{n_{R\alpha}}$) where the sum runs over all possible states
$\alpha$.  For a bosonic sector $ f(z) = \eta^{-24}(z)= {1\o
z}f(z)^{24}\ $ (the extra factor $1\o z$ corresponds to the tachyon),
while for a superstring sector $f(z) = {\vartheta_2^4(z)\over
\eta^{12}(z)}$. As before we compute the asymptotic level densities
using a saddle point evaluation
\eqn\dnev{
d(n) = {1\over2\pi i} \oint {dz\over z^{n+1}} f(z)\, .  }
The resulting level densities for left and right sectors are
\eqn\dnrs{
d(n_L)\sim n_L^{-(d+1)/4} e^{ 4\pi \sqrt{a_L n_L}}\ ,\ d(n_R)\sim
n_R^{-(d+1)/4} e^{ 4\pi \sqrt{a_R n_R}} }
where $a_{L,R}=\hf$ for a superstring sector and $a_{L,R}=1$ for a
bosonic string sector (so for the heterotic string $a_L=1$ and
$a_R=\hf$).  Physical states must also obey the level-matching
condition ${
\rm L}_0 -{\bar
{\rm L}}_0 =0$, or $n_L -n_R = -\sum_i m_i n_i$.  This and \compen\
give $$
\eqalign{ 8n_L& = E^2 - k^2-\left({n_i\over R_i} + 2R_im_i\right)^2\cr
          8n_R& = E^2 - k^2-\left({n_i\over R_i} - 2R_im_i\right)^2\ .
} $$ Therefore the single-string density of states is (including the
chemical potentials $\kappa$, $\mu$ and $\nu$)
\eqn\sstate{\eqalign{ \omega(E,\kappa,\mu,\nu)dE \sim V_D\int&
{d^Dk\over (2\pi)^D}\sum_{m_i,n_i} {EdE\over4} e^{2\pi i (\kappa\cdot
k + \mu\cdot m +\nu\cdot n)}
\cr
&{e^{\left\{ 4\pi\sqrt{a_L n_L(E,k,n_i,m_i)} + 4\pi\sqrt{a_R
n_R(E,k,n_i,m_i)} \right\}} \over n_L(E,k,n_i,m_i)^{(d+1)/4}
n_R(E,k,n_i,m_i)^{(d+1)/4} }
\ . }}
Here $D=d-c-1$ is the number of non-compact space directions, and
$V_D$ is the volume of the non-compact directions.)

Note that for each $R_i$ there are regions with different behaviour.
If $R_i \gg E \gg 1/R_i$ we ignore the winding quantum numbers $m_i$
and the sum over the closely spaced momentum levels is well
approximated by an integral over a continuous spectrum of momentum;
thus the string behaves as if the direction $i$ were noncompact.
Likewise, if $1/R_i
\gg E\gg R_i$ the sum over momenta drops out and the sum over winding
becomes continuous; this situation is dual to the previous one under
$R\rightarrow 1/2R$.  Finally we have the truly high-energy regime, $
E \gg R_i, 1/R_i$.  Here we can closely approximate the sum over $m_i,
n_i$ by an integral $dm_i dn_i$.  Furthermore, for large $E$ we can
evaluate the integral \sstate\ by saddle point methods.  If we expand
the arguments of the square roots to first order in $k^2$, $(n_i/R_i
\pm 2Rm_i)^2$ (valid for large $E$ in the saddle point approximation
since the region $(n_i/R_i \pm 2Rm_i)^2/E^2 \ll 1$, $k^2/E^2 \ll 1$
dominates), the integral becomes quadratic and yields
\eqn\sdens{\omega(E,\kappa,\mu,\nu) \sim {1\over E^{D/2+1}}
e^{\beta_H(\kappa,\mu,\nu) E}}
where
\eqn\vhag{\beta_H(\kappa,\mu,\nu) = \beta_{H}-
\sqrt{2}\pi \left\{ {\kappa^2\over\sqrt{a_L} + \sqrt{a_R} }  +
{ \left({\mu\over 4R} + {\nu R\over2} \right)^2\over \sqrt{a_L}} +
{\left({\mu\over 4R} - {\nu R\over2} \right)^2\over\sqrt{a_R}}
\right\}}
and $$
\beta_{H} = \sqrt{2} \pi \left(\sqrt{a_L}
+\sqrt{a_R} \right) $$ is the usual Hagedorn temperature \bowgi .  Thus in
effect the Hagedorn temperature depends on the chemical potentials.

This result can now be used to derive the multi-string density of
states as sketched in the previous section.  First we will compute the
Maxwell-Boltzmann contribution \boltz\ to the total density of states,
and then we will argue that the corrections due to Bose-Einstein or
Fermi-Dirac statistics do not substantially alter the asymptotic form
of the density of states.

The MB expression \boltz\ becomes a sum of terms of the form
\eqn\boltzdens{ {1\over n!} \int\prod_{i=1}^n
{ dE_i\over E_i^{D/2+1} } e^{\beta_H(\kappa,\mu,\nu) E_i}
\delta\left(E-\sum_{i} E_i\right) \ .}
The integrals over $E_i$ are divergent at the lower end and must be
cut off at some energy $m_0\roughly>M_s$ where the asymptotic
expression \sdens\ is no longer valid.  It is clear that the dominant
contribution to \boltzdens\ is that in which most of the energy is in
one string, $E_i\approx E$.  Thus the integral is approximately
\eqn\mndens{ \eqalign{ \Omega_n(E,\kappa,\mu,\nu) \sim &{1\over(n-1)!}
{e^{\beta_H(\kappa,\mu,\nu) E}\over E^{D/2+1}} \left({2\over
Dm_0^{D/2}}\right)^{n-1}\ {\rm for}\ D> 0;\cr
\sim & {1\over(n-1)!}
{e^{\beta_H(\kappa,\mu,\nu) E}\over E} \left[ \ln(E/m_0)\right]^{n-1}\
{\rm for}\ D=0.}}

Summing over $n$ then gives
\eqn\mnii{\eqalign{ \Omega(E, \kappa,\mu,\nu) &\sim
{e^{\beta_H(\kappa,\mu,\nu) E}\over E^{D/2+1}}e^{2/\left(Dm_0^{D/2}
\right)}\ {\rm
for}\ D>0;\cr &\sim {1\over m_0} e^{\beta_H(\kappa,\mu,\nu) E} \ {\rm
for}\ D=0.}}
The density of states without the conservation constraints is given by
setting $\kappa=\mu=\nu=0$; this expression has been found in the
literature.  To obtain the density of states including the constraint
of zero total momentum and winding, we use \vhag\ and integrate over
$\kappa,\mu,\nu$ to find
\eqn\dens{\ome \sim {e^{\beta_H E}\over E^{d-\delta_D}}}
where we define $\delta_D=1$ if $D=0$ and $\delta_D=0$ otherwise.
Imposing the constraints changes the power of $E$; this is in contrast
to other thermodynamic situations where conservation laws are not so
important essentially because one string carries most of the energy
and bears the burden of the conservation laws.

We can also compute corrections to
\dens\ due to Bose-Einstein and Fermi-Dirac statistics.
One finds that the effect of these corrections is to multiply \dens\
by a function that depends on low-energy physics but to leading order
not on $E$.

Thus for the type II string the asymptotic density of states is
\eqn\iidens{\ome \sim {e^{\beta_H E}\over E^{10-\delta_D}}\ .}
If we consider the heterotic $E_8\times E_8$ or ${\rm Spin}(32)/Z_2$
string we see that conservation of the 16 U(1) charges should also be
imposed.  In the bosonic formulation this is equivalent to
momentum/winding conservation for the internal degrees of freedom,
and, as one can easily convince oneself by a slight modification of
the above argument, gives an extra factor of $1/E^8$.  Thus the
asymptotic density of states for the heterotic string is
\eqn\hetdens{\ome \sim {e^{\beta_H E}\over E^{18-\delta_D}}\ .}

It will be useful for us to deduce which string configurations
contribute most significantly to the total string density of states;
this can be done by inspection of \boltzdens\ or the corresponding
expression including quantum statistics.  As we have already noted,
the dominant configuration for the integral for a fixed number $n$ of
strings is when $E_i\approx E$, \ie\ when most of the energy is in
{\it a single string}.  For $D>0$ it also appears that string
configurations with a small total number of strings dominate the sum,
but for $D=0$ configurations with a large total number of strings
dominate the sum. (These qualitative features also hold true when
corrections due to quantum statistics are included.)  We will return
to a discussion of this single string dominance of the energy in
section 7
\bigskip
\centerline{\bf 7. Path-Integral Derivation}\bigskip

In the meantime, we turn to the second derivation of the string density
of
states \nref\polch{
J. Polchinski, {\sl Commun. Math. Phys.} {\bf 104} (1986)
37.}\nref\mcl{ B. McClain and B. D. B. Roth, {\sl Commun. Math. Phys.}
{\bf 111} (1987) 539.}\nref\brta{ K. H. O' Brien and C. I. Tan, {\sl
Phys. Rev.} {\bf D36} (1987) 1184.}\nref\bowrev{M. J. Bowick,
In Proceedings of the 1989 Summer School In High Energy Physics and
Cosmology, in the {\sl ICTP Series in Theoretical Physics} {\bf 6},
eds. J. C. Pati, S. Randjbar-Daemi, E. Sezgin and Q. Shafi, World
Scientific, 1990.}\refs{\polch {--} \bowrev} and \bowgi .
Once again \densen\ is used to relate
$\ome$ to the complex function $\zbe$.  Next recall that
the free energy $F=-{1\over\beta}\ln Z$ for a single free
particle of mass $M$
can be computed from a first-quantized one-loop path
integral on a space with
compactified time direction of circumference $\beta$.  The
action is $S_p =\hf\int_0^1 d\tau [e^{-1}
(dX/d\tau)^2 + e M^2]$, and in the
path integral we integrate over all maps $X^\mu(\tau)$
of the circle into the target spacetime $S^1 \times R^{d-1}$ and over
all one-metrics $g=e^2$ on the circle.   Explicitly,
\eqn\pathi{2\ln Z =
\sum_{n=-\infty}^{\infty} \int {\cald g\over {\rm Vol(Diff)}}
\int_n \cald X e^{-S_p} (-1)^{n {\hat F}}}
where in the path integral over metrics we must eliminate the volume
of the diffeomorphism group, and where the boundary condition on the
$X$ path integral is $X^0(1)=X^0(0) + n\beta$, $X^i(1) = X^i(0)$;
$n$ is the number of times that the circle winds around the compact
time direction.
We also define ${\hat F}$ to be zero for bosons and one for fermions;
the extra factor $(-1)^{n{\hat F}}$ corresponds to anti-periodic boundary
conditions for fermions in the second-quantized formalism.
One easily shows (taking into account the Killing vector on the circle)
that
$$
\int {\cald g\over {\rm Vol(Diff)}} = \int_0^\infty {ds\over s}
$$
where $s$ is the proper length of the circle.  One can also show
$$
\int_n\cald X e^{-S_p} = (V\beta)(2\pi s)^{-d/2}
e^{-{n^2\beta^2\over2s}
-{M^2 s\over2}}
$$
where $V$ is the spatial volume.
These give the free energy
\eqn\free{F = -{V\over2}
\int_0^\infty {ds\over s}
(2\pi s)^{-d/2}
\sum_{n=-\infty}^{\infty} e^{ - {n^2\beta^2
\over 2s} -{M^2s\over2}} (-1)^{n{\hat F}}\ . }
It is fairly simple to demonstrate that this is equivalent to the
standard expressions for the free energy of a single free particle.

Likewise, the
free energy for string is obtained by doing the functional integral
over maps of the torus into $S^1\times R^{d-1}$, with the (conformal
gauge-fixed) action which we may
take to be that of the heterotic string (our units are such that $\alpha'
= \hf$)
$$
S= {1\over2\pi} \int d^2\sigma \left(\del_z X^\mu \del_\zb X_\mu + {\bar
\psi}^\mu \del_z {\bar \psi_\mu} + \psi^\mu\del_\zb \psi_\mu +\lambda^i\del_\zb
\lambda^i\right)\ .
$$
(The $\lambda_i$ provide a fermionic representation of
the gauge degrees of
freedom.)  The maps are allowed to wind in the time direction on the torus,
but not in the space direction, as in the case of the particle.
The result is
\eqn\sfree{\eqalign{ F= -{V\over4}\pi^{-5} \int_0^\infty
{d \tau_2\over \tau_2}& \int_{-\hf}^{\hf}
d\tau_1 (2\pi\tau_2)^{-5} {1\over\eta(q) ^{8}{\bar \eta(\qb)}^{12}}
ch[{\widehat G}](q)\cr & \sum_{\pi_1,\pi_2 =0,1}
\Cb_{\pi_1\pi_2}(\qb) \sum_n exp\left\{-
{\beta^2n^2\over 2\pi\tau_2}\right\}
(-1)^{n\pi_1}\ .}}
Here $q=\exp\{2\pi i(\tau_1+i\tau_2)\}$,
$\eta$ is the Dedekind function; $ch[{\widehat G}]$ is the character
function of the gauge group
$G= E_8\times E_8$ or SO(32), $ch[{\widehat E_8 \times \widehat E_8}] =
(\vartheta_2^8 + \vartheta_3^8 + \vartheta_4^8)^2/4\eta^{16}$,
$ch[{\widehat{SO(32)}}]=
(\vartheta_2^{16}+\vartheta_3^{16}+\vartheta_4^{16})/2\eta^{16}$;
and
${\Cb}_{00} = \varthb_3^4$, ${\bar C}_{01} = - \varthb^4_4$,
${\bar C}_{10}
= - \varthb_2^4$, and $\Cb_{11}=0$.
This expression is analogous to \free; the integral over $\pi\tau_2$
corresponds to that over $s$, the sum over $n$ is the sum over sectors
where the time direction on the torus wraps $n$ times around the target
time dimension, and the rest of the expression integrated over $\tau_1$
gives the sum over all single string states of
$(-1)^{n{\hat F}}\exp\{-M^2\pi\tau_2/2\}$.
We could therefore alternatively
derive this expression directly from \free\ by thinking of the
string as a collection of particles corresponding to its various modes.

To make this expression look more familiar from the string point of
view, we recall that the region $\cals$ of the $\tau=\tau_1 +i\tau_2$ plane
$|\tau_1|<\hf$, $0<\tau_2<\infty$ corresponds to an infinite number
of copies of the fundamental domain $\calf$ defined by $|\tau_1|<\hf$,
$|\tau|>1$; $\cals$ is tiled by images of $\calf$ under maps $\tau'
= (p\tau+q)/(r\tau+s)$, $p,q,r,s\in {\bf Z}$,
$ps-qr=1$.  The free energy
for the heterotic string
can therefore be rewritten  as an integral over the
fundamental domain
\eqn\ffree{\eqalign{F=-{V\over4}\pi^{-5}&
\int_\calf {d^2 \tau\over \tau_2}
(2\pi\tau_2)^{-5} {1\over\eta(q) ^{8}{\bar\eta(\qb)}^{12}}
ch[{\widehat G}](q) \sum_{\pi_1,\pi_2 =0,1} {\bar
C}_{\pi_1\pi_2}(\qb)\cr & \sum_{m,n}
exp\left\{- {\beta^2\over 2\pi\tau_2}\left(
m^2\tau_2^2 +(n-m\tau_1)^2\right)\right\}
(-1)^{m\pi_2 + n\pi_1 +mn}}}
and likewise for the type II string
\eqn\iifree{\eqalign{F=-&{V\over8}\pi^{-5}
\int_\calf {d^2 \tau\over \tau_2}
(2\pi\tau_2)^{-5} {1\over\eta(q) ^{8}{\bar\eta(\qb)}^{8}}
\sum_{\pi_1,\pi_2 =0,1} \sum_{{\bar \pi}_1,{\bar \pi}_2 =0,1}
C_{\pi_1\pi_2}(q)
{\bar
C}_{{\bar \pi}_1{\bar \pi}_2}(\qb)\cr & \sum_{m,n}
exp\left\{-{\beta^2\over 2\pi\tau_2}\left(
m^2\tau_2^2 +(n-m\tau_1)^2\right)\right\}
(-1)^{m(\pi_2 +{\bar\pi}_2) + n(\pi_1+{\bar\pi}_1)}\ .}}
{}From \ffree\ one can proceed to calculate the asymptotic density of
states .

In these expressions
the sum over $m$ is the sum over maps of the torus in which the
{\it space} direction of the torus is wound $m$ times about the target
$S^1$.  While one might think of these as corresponding to
strange states of the string
in which it wraps around the time direction, this interpretation is
not necessarily correct if the fundamental expression for the free energy
is \sfree\ which directly corresponds to
$Z(\beta) = tr\, e^{-\beta H}$.  This
trace contains only ordinary string states and not the strange `winding
states;' the latter appear to arise only as mathematical artifacts when
we transform the expression for $\ln Z$ as an integral over $\cals$
into an integral over $\calf$.

Note the following interesting point.  \sfree\
is equivalent to the modular invariant \ffree\ the cosmological constant
$\Lambda=0$.  However, if $\Lambda\neq 0$, \sfree\ gives $\infty\cdot
\Lambda + F\,{}'$ and \ffree\ gives $\Lambda+ F\,{}'$ where $\Lambda$ is
cosmological constant as it is usually defined in string theory (as
an integral over $\calf$) and $F\,{}'$ is the temperature dependent part
of $F$. This is also a difficulty for string field theory -i.e. to
naturally produce the fundamental domain.
\bigskip
\centerline{\bf 8. Physical Behaviour of Strings at High Temperature}
\bigskip

We now turn to an attempt to extract a better understanding of the physics
of strings at high temperature from the above calculations
\refs{\bowgi , \bowrev}.  To start
with, assume that we have at our disposal a thermal bath that can interact
with the string ensemble to heat it up.  Within the context of fundamental
strings this assumption is not necessarily valid as all matter is
presumably made out of strings, but we make this initial
assumption
to aid us in isolating the source of any strange behavior that
we may encounter; we will see that
the assumption of the existence of such a reservoir
doesn't obviously appear to be related to any such behavior.
Therefore suppose that we have a heat reservoir with
a density of states $\Omega_r(E)$.

The total density of states for
the string ensemble in contact with this reservoir is
$$
\Omega_T(E) = \int_0^E dE_s \Omega(E_s) \Omega_r(E-E_s)\ .
$$
If we assume that the reservoir is large (in a sense to be made precise
shortly), then most of the energy will be in the reservoir, \ie\ $E\gg
E_s$.
In this case we can expand
$$
\ln\Omega_r(E-E_s)\cong \ln\Omega_r(E) -
E_s {\del\over\del E} \ln\Omega_r(E)
+\hf E_s^2 {\del^2\over\del E^2} \ln\Omega_r(E) +-\cdots\ .
$$
We identify $ {\del\over\del E} \ln\Omega_r(E) = \beta$ as the inverse
temperature of the reservoir.  We can drop the higher terms when the
reservoir is sufficiently large; precisely,
$
E_s^2{\partial ^2\o \partial E^2}S_r(E)\ll 1
$ or
\eqn\bigres{{1\over\beta^2}
C_{V{\scriptscriptstyle r}}=- {\del E\over \del \beta}
\gg E_s^2\ .}
Thus we can use the ensemble
\eqn\omt{\Omega_T(E) = \Omega_r(E) \int_0^\infty dE_s e^{-\beta E_s}
\Omega(E_s)}
as long as we ensure that the heat capacity of the reservoir
is large enough in the sense of \bigres; replacing the upper limit by
$\infty$ has no significant effect.

Now we are prepared to study the properties of the string gas.  As we
raise the temperature, \omt\ stays well defined all the way to $\beta_H$.
Even at $\beta_H$ \omt\ is well defined because of the power law tail
$\sim 1/E^{(d-\delta_D)}$.  (Of course $\ome$ takes a different, convergent,
form for low energy; since the string is a gas of massless particles
at very low energies, we expect $\ome\sim \exp\left\{a{D+1\over D}
E^{D/(D+1)}
V_D^{1/(D+1)}\right\}$ there for some constant $a$).
Furthermore, the mean energy density
\eqn\meane{ \langle {\cal E}
\rangle = {\int_0^\infty dE e^{-\beta E} {\cal E} \ome\over
\int_0^\infty dE e^{-\beta E} \ome } }
is finite for $\beta=\beta_H$; $e^{-\beta_H E} \ome$ will presumably look
something like Fig.~1 where the maximum $E_0$ is somewhere between the
low energy and high energy domains where $\Omega(E)$ is known, \ie\
${\cal E}_0$ will be of order one in string scale units.

What happens when we attempt to raise the temperature further?  It is
apparent from \omt\ that for $\beta<\beta_H$ this ensemble is not defined;
its mean energy becomes infinite.  Therefore it seems that we cannot
pass $\beta_H$.  But in principle we could imagine adding more energy
to the string system in an attempt to raise its temperature; where does
this energy go?  A clue was pointed out at the end of section five:
at large energies the density of states is dominated by one (or a small
number of) highly energetic, and therefore long and wiggly,
strings.  Thus it seems quite reasonable that any energy that we add
to the system in an attempt to raise the temperature goes into making
strings in very high oscillation modes.
The low mass modes are populated with a thermal distribution
at $\beta_H$, and the rest of the energy is in oscillation modes.
It may be useful to think of this transition to long-string dominance
as a higher-order
phase transition with
an infinite latent heat.  The low temperature phase is what we know,
and the `high temperature phase' is the configuration where the fractional
energy in long string is one, but this phase
is inaccessible because it corresponds
to infinite energy. (This is assuming that the fundamental degrees of
freedom are not drastically modified at some finite energy density,
as could well be the case.)
\bigskip
\centerline{\bf 9. X-Y Model and the Berezinski\u i-Kosterlitz-Thouless
Transition}\bigskip

Consider the string action in the conformal gauge
\eqn\cgact{
S={1\o 2\pi }\int d\sigma d\tau\ (\partial _\alpha X^\mu \partial ^\alpha
X^\mu )\, ,
}
where we take $\alpha ^{\prime }=1/2$. Compactifying in the time
direction, in order to account for the finite temperature,
and using periodic boundary
conditions for closed string
$X^\mu(\sigma +\pi ,\tau )=X^\mu(\sigma ,\tau)$
and $X^\mu(\sigma ,\tau+\pi )=X^\mu(\sigma ,\tau)+n\b \delta^\mu_0$,
we obtain the mode expansion for the time coordinate
\eqn\tmex{
X^0(\sigma ,\tau )=\tilde X^0+{2\pi n\tau \o\beta} +{m\beta \sigma \o
\pi} +\hbox{ \rm(oscillators)}
}
and
\eqn\enper{
{1\o 4}E^2=N_L+\frac 12\left( \frac{n\pi }\beta +\frac{m\beta }{2\pi }%
\right) ^2-1+N_R+\frac 12\left( \frac{n\pi }\beta -\frac{m\beta }{2\pi }%
\right) ^2-1
}
with the constraint $N_L-N_R=m\cdot n$. The state $N_L=N_R=0$ with
$n=0$ and $m=\pm1$ has
\eqn\mden{
E(T)^2=M(T)^2=-8+{1\o\pi^2T^2}
}
For $T<T_H=1/2\pi \sqrt{2}$ its mass-squared is positive but at
$T=T_H$ it becomes massless and for $T>T_H$ tachyonic.
Call the $m=\pm 1$ winding states $\varphi $ and $\varphi ^{*}$ respectively.
Thus the
Hagedorn divergence $(T>T_H)$ corresponds to the existence of a tachyon and
a divergence at the boundary $\tau _2\rightarrow \infty $ of the moduli
space in the path-integral language \refs{\sath {--} \kog}
\eqn\divmod{
F(\beta )\sim (const.)\int\limits^\infty d\tau _2\ e^{-M^2(\beta )\tau
_2}\rightarrow \infty \quad\hbox{ \rm for }\quad M^2(\beta )<0\, .
}
The action for the $\varphi $ mode given by
$X^0=\frac \beta {2\pi }\varphi$ with
$\varphi(\s, \t+\pi)=\varphi(\s ,\t)+2\pi$ is
\eqn\vilact{
\eqalign{
S_\varphi &=\frac{\beta ^2}{8\pi ^3}\int (\pa\varphi )^2\cr
          &=\frac g2\int (\pa\varphi )^2\, ,\qquad
               g\equiv \frac{\beta ^2}{4\pi ^{3}}\, .\cr
        }
}
The action $S_\varphi$ is that of the Villain model which is the
continuum limit of that of the X-Y model (planar magnet)
\eqn\xyact{
S=-\frac g2\sum\limits_{<i,\delta >}\cos (\phi _i-\phi _{i+\delta })\, .
}
The vortex solutions of the model are given by
\eqn\vrtsl{
\varphi (z)=-\frac i2m\ \ln \left[ \frac{z-z_0}{\bar z-\bar
z_0}\right]\, ,
}
where $z_0$ is the position of the vortex and $m$ is the winding
number. The corresponding free energy is
\eqn\frvrt{
F_m=(\frac{\pi g}2m^2-1)\ \ln\left( \frac A{a^2}\right)\, ,
}
where $a$ is the lattice spacing cutoff.
The above model has an infinite order Berezinski\u i-Kosterlitz-Thouless
phase transition at $g_c={2\o\pi}$ or $\b_c=2\sqrt{2}\pi=\b_H$!
In the low temperature regime $(g>g_c)$ the system is dominated by
spin wave excitations and is in an unscreened scale invariant phase
with power law spin-spin correlations and tightly bound
vortex-antivortex pairs. In the high  temperature regime $(g<g_c)$
the topological order is destroyed, the vortices unbind and are free
(vortex condensation). The phase is screened with exponentially
decaying spin-spin correlations characterized by a coherence length
$\xi$ which breaks the scale invariance.

Consider a chemical potential $\mu$ for vortices, where $\mu$
corresponds to the energy of dissociation of a vortex-antivortex bound
state. The RG flows for the system are in the fugacity
$y$ $(y=e^{-\mu /kT})$ and $T$ plane as shown in Fig.1.

For $T<T_c$ and $y$ small the system flows to $y=0$
or to $ \mu =\infty $
and vortices are bound (region I). For $T>T_c$ and $y$ small the system
flows off to $y$ large out of the domain of perturbation theory but $\mu $
is effectively small and vortices can unbind.

To describe string theory in this phase one would have to look at the
coupled $\beta $-functions for all the massless moduli at $\beta =\beta
_H $ which include $g_{\mu \nu }, B_{\mu \nu }$, the dilaton
$\sigma $
and $\varphi $. This might have solutions corresponding to a conformal fixed
point for $T>T_H$. Atick and Witten \atwt\ suggested that
$\varphi $ gets a v.e.v. $<\varphi >$ before $T_H$ thus obviating
the difficulties of the Hagedorn divergence.
But they found the effective potential for $\varphi $ obtained from its
interactions with the dilaton $\sigma $ is of the type
$$
V_{eff}(\varphi )\sim m^2(T)\varphi ^{*}\varphi -\hat \lambda
(\varphi ^{*}\varphi )^2
$$
with
$$
\hat \lambda\ >0\quad{\rm and}\quad m^2(T)=-8+1/\pi ^2T^2\, .
$$
This implies a first-order phase transition. $V_{eff}(\varphi )$ must, however,
be stabilized by higher-order terms and $<\varphi >$ is not calculable in
perturbation theory.

Atick and Witten argued that, had $\hat \lambda $ been negative, there would
be a {\it second order} phase transition at $T=T_H$. For $T$ just
greater than $T_H$ then%
$$
\left| <\varphi >\right| ^2=\frac{\left| m^2\right| }{2\hat \lambda }
$$
and the free energy is

$$
F=V(<\varphi >)=\frac{-m^2}{4\hat \lambda }=\frac{-m^2}{4\lambda g^2T}
$$
This is a genus zero (sphere) contribution to the free energy and implies
the world-sheet is spontaneously tearing to become non-simply-connected.
They then argue that this picture is valid even for the first-order
transition case $\hat \lambda >0$. This is not, so far, justified by any
calculation and fails in the weak coupling limit $g\rightarrow 0$.
\bigskip
\centerline{\bf Appendix}\bigskip

In this appendix we compute the asymptotic form of the level density
$p(N)$ which led to \aspn . From \contlim\ we have that
\eqn\strt{
p(N)=\frac 1{2\pi i}\oint dz\left( \frac{-\ln z}{2\pi }\right) ^{1/2}\exp
[g(z)]\, ,
}
with $g(z)=\frac{-\pi ^2}{6\ell nz}-(N+c)\ln z$ and $c=23/24$. Then
$g^{\prime }(z)=\frac{\pi ^2}{6(\ln z)^2}\cdot \frac 1z-\frac{N+c}z$,
which has a maximum at $\ln z_0=-\frac \pi {\sqrt{6}}\cdot \frac 1
{\sqrt{N+c}}$. Then
\eqn\rslt{
g(z_0)=\frac{2\pi }{\sqrt{6}}\sqrt{N+c}\, .
}
Thus the leading behavior of $p(N)$ is
\eqn\aspt{
\eqalign{
p(N)&\sim \exp \left( \pi \sqrt{\frac
      23}\sqrt{N+\frac{23}{24}}\right)\cr
    &\sim \exp \left( \beta _c\sqrt{N}\right)\cr
        }
}
with $\beta _c=\pi \sqrt{\frac 23}$. Corrections are obtained by
looking at the quadratic fluctuations
\eqn\quadr{
g(z)=g(z_0)+\frac 12(z-z_0)^2g^{\prime \prime }(z_0)+\ldots\, ,
}
where $ g^{\prime \prime }(z_0)=\frac{2\sqrt{6}}\pi
(N+c)^{3/2}\exp \left( \pi \sqrt{\frac 23}\frac 1{\sqrt{N+c}}\right)$.
Then
\strt\ is a Gaussian integral which gives
\eqn\fnlpn{
p(N)=\frac 1{4\sqrt{3}}\frac 1N\exp \left( \pi \sqrt{\frac 23}\sqrt{N}
\right)\, .
}
For the string in $d$ dimensions \fnlpn\ easily generalizes to
\eqn\dimsp{
p(N)=(const.)N^{-(d+1)/4}\exp \left( \pi \sqrt{\frac
23(d-2)}\sqrt{N}\right)\, .
}
%


\bigskip\bigskip
\bigskip
\centerline{\bf Acknowledgements}
\bigskip

This
was supported by the Outstanding Junior Investigator Grant DOE
DE-FG02-85ER40231. I also wish to thank Steve Giddings for
collaboration on which part of this review is based and Konstantinos
Anagnostopoulos for much assistance in the preparation of this manuscript.

\listrefs
\eject
\figures
\fig1{
R.G. flows for the Villain model. The $x$ axis is $T_c/T-1$ and the y
axis the fugacity  $e^{-\mu/kT}$. The $x$ axis is a line of trivial
fixed points that can be reached from region I. Points from region II
flow to large y and out of the domain of validity of perturbation
theory.
}
\bye